%% file: main.tex
\documentclass{article}
\usepackage{microtype}
\usepackage{graphicx}
\usepackage{subcaption}
\usepackage{booktabs}
\usepackage{amsmath,amssymb,amsfonts}
\usepackage{textcomp}
\usepackage{xcolor}
\definecolor{revisionplaceholder}{HTML}{B01872}

\usepackage[hypertexnames=false]{hyperref}

\usepackage[accepted]{mlsys2025}

\makeatletter
\renewcommand{\Notice@String}{}
\makeatother

\hypersetup{pdfsubject={Distributed LLM inference}} %

\mlsystitlerunning{Decode Token Parallelism for Scalable LLM Decoding}
\begin{document}
\twocolumn[
\mlsystitle{Decode Token Parallelism: Memory-Efficient Distributed Attention Parallelism for Scalable LLM Decoding}
\begin{mlsysauthorlist}
\mlsysauthor{Susav Shrestha}{nvsc}
\mlsysauthor{Brad Settlemyer}{nvsc}
\mlsysauthor{Narasimha Reddy}{tamu}
\end{mlsysauthorlist}
\mlsysaffiliation{nvsc}{NVIDIA, Santa Clara, United States}
\mlsysaffiliation{tamu}{Texas A\&M University, College Station, United States}
\mlsyscorrespondingauthor{Susav Shrestha}{sshrestha@nvidia.com}
\mlsyskeywords{Large Language Models, Transformer, Token Parallelism, Distributed Systems, Parallel Computing, Attention Mechanism, KV Cache, Model Parallelism}
\vskip 0.3in
\begin{abstract}

Serving large language models at high concurrency with long contexts requires KV-cache capacity and attention bandwidth that grow with the decode pool. Existing parallelization strategies trade one against the other: data parallelism replicates model weights, limiting KV capacity; tensor parallelism stops scaling attention at the model's KV-head count and pushes dense collectives across nodes; and decode context parallelism (DCP) recovers capacity by sharding the sequence but adds cross-node aggregation of partial attention results. We observe that during decode, dense layers and attention scale differently: dense layers amortize weight reads across the batch and run efficiently within a single NVLink domain, whereas attention cost grows with every active request's KV cache. We present \textbf{Decode Token Parallelism (TKNP)}, a decode-pool architecture that decouples weight placement from attention ownership. TKNP distributes a single model copy across nodes in balanced layer blocks. The active node holding a block executes its projections and MLP for the full batch using intra-node tensor parallelism, while every node computes attention for its own requests against locally resident KV cache. This extends request-local attention across nodes without replicating weights: only current-token activations cross the network, dense all-reduces remain within NVLink, and attention requires no cross-node merging of partial results. Implemented in vLLM and evaluated on Llama~3.1~405B and Nemotron-4~340B on up to 64 H100 GPUs, TKNP improves completion throughput for long-context, high-concurrency workloads by \textbf{up to 1.7$\times$ over DCP} and by \textbf{up to 2.06$\times$ over data parallelism} at matched GPU count.

\end{abstract}
]
\printAffiliationsAndNotice{}

\input{Paper/1.Introduction}

\input{Paper/2.Background}
\input{Paper/3.Motivation}
\input{Paper/4.TokenParallel}

\input{Paper/5.Evaluation}
\input{Paper/6.FutureWork}
\input{Paper/7.Conclusion}

\bibliographystyle{mlsys2025}
\bibliography{references}

\end{document}

%% file: Paper/1.Introduction.tex
\section{Introduction}

\begin{figure}[!b]
    \centering
    \includegraphics[width=\linewidth]{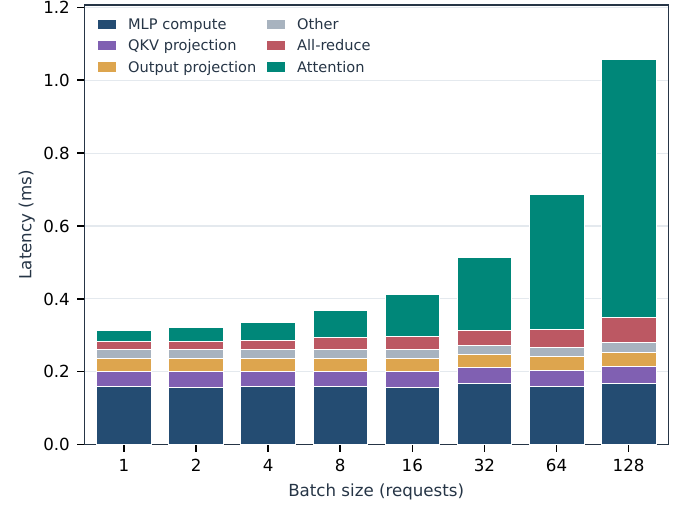}
    \caption{\textbf{Decode-stage latency breakdown for a single decoder layer of Llama 3.1 405B} on eight H100 GPUs (TP8) with 32K context. Attention grows with batch size, accounting for nearly 70\% of summed component latency at batch 128.}
    \label{fig:TknpMotivation}
\end{figure}

\begin{figure*}[!t]
    \centering
    \includegraphics[width=0.88\textwidth]{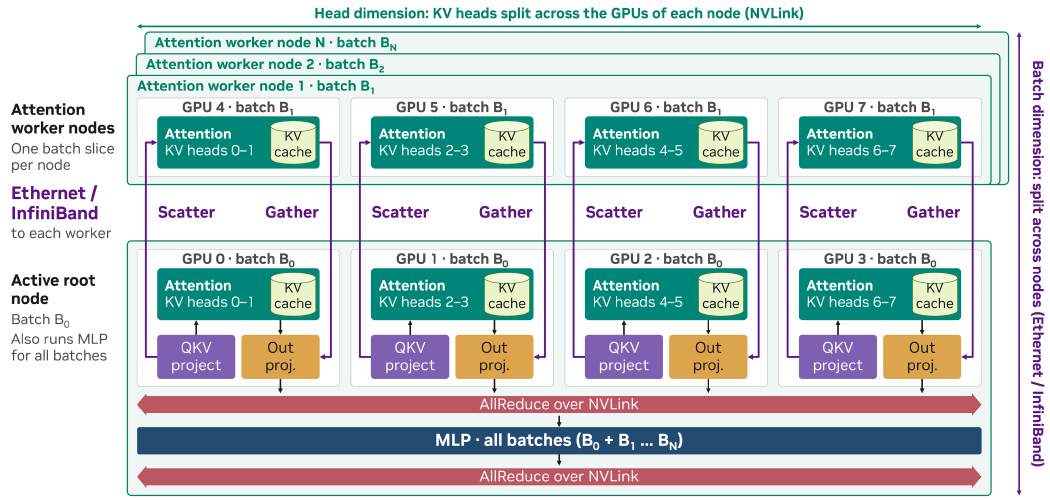}
    \caption{\textbf{Token Parallel architecture within a layer block.}
    TP partitions KV heads within each node; TKNP partitions requests across
    nodes. The active root executes QKV and output projections and the MLP
    for the full batch, while all nodes compute attention for their own
    requests using local KV cache. Scatter/gather exchanges current-token
    activations over Ethernet or InfiniBand; dense all-reduces remain within
    NVLink. Four GPUs per node are shown for illustration. Weight storage
    on attention workers is omitted; the active root changes with the block.}
    \label{fig:TokenParallel}
\end{figure*}

Large language model serving must sustain high throughput while responding promptly to many concurrent requests. Modern inference engines use continuous batching to amortize model-weight reads across requests and admit new work as earlier requests finish~\citep{orca,vLLM,sglang}. For models with hundreds of billions of parameters, however, the memory required to store model weights leaves limited space for long-context KV caches. Increasing the batch improves the efficiency of dense computation, but also increases KV-cache demand, preventing additional requests from becoming resident.

The root of this tension is that dense layers and attention scale differently during decode.
MLPs and projections process only newly generated tokens, allowing weight reads to be shared across the batch, whereas attention reads the accumulated KV cache of every active request. As Figure~\ref{fig:TknpMotivation} shows, attention approaches 70\% of single-layer decode latency for Llama 3.1 405B at batch 128 with 32K context while MLP and projection latency remain relatively stable. Scaling such workloads therefore requires additional attention bandwidth and KV capacity while preserving efficient dense execution. On multi-node GPU clusters, this must also account for the interconnect hierarchy: GPUs communicate through high-bandwidth NVLink within a node, while cross-node communication uses Ethernet or InfiniBand.

Existing parallelization strategies address these requirements through different tradeoffs. Data parallelism (DP) keeps execution local but replicates model weights, shrinking aggregate KV capacity. Tensor parallelism (TP) replicates KV heads once its degree exceeds the KV-head count~\citep{megatronlm,GQA} and moves dense all-reduces onto the inter-node network when it spans nodes. Decode context parallelism (DCP) shards KV cache beyond this limit but must combine partial attention results across ranks. Pipeline parallelism (PP) avoids replication but confines each layer's attention to one stage and repeats dense weight reads across microbatches~\citep{gpipe,pipedream}. Data-parallel attention for MoE serving~\citep{sglang2024dpattention} keeps attention local to each request's KV owner but replicates attention weights on every rank, while attention--FFN disaggregation~\citep{attentionoffloading,megascaleinfer} leaves the weight pool's spare memory unused for KV cache. These tradeoffs motivate separating the placement of model weights from the ownership of attention and KV cache.

We introduce \textbf{Decode Token Parallelism (TKNP)}, a decode-pool architecture for prefill--decode disaggregated serving that \textbf{decouples where model weights reside from where attention is computed}.
TKNP distributes one model copy in stationary layer blocks, balancing weight storage to leave similar KV-cache budgets across nodes. Each layer's projections and MLP execute on its weight-owning active root node using local TP, while all nodes compute attention for their assigned requests. As shown in Figure~\ref{fig:TokenParallel}, attention scales across heads within nodes and requests across nodes, 
with only current-token activations crossing the network, dense all-reduces staying within NVLink, and no cross-node merging of partial attention results.

We implement TKNP in vLLM and evaluate it against DP, TP, PP, and DCP baselines on Llama~3.1~405B \cite{llama3} and Nemotron-4~340B \cite{nemotron4} using up to 64 H100 GPUs. Our contributions are:

\begin{itemize}
    \item We separate attention ownership from weight placement, enabling \textbf{efficient attention scaling} across nodes without replicating model weights.
    \item We balance weight placement so that \textbf{every node contributes approximately equal KV capacity}, matching sequence-sharded layouts and providing up to \textbf{5.0$\times$ that of DP} on 64 GPUs.
    \item We propose dual-active-root microbatching to \textbf{overlap dense computation and attention}, improving throughput by up to 11.1\%.
    \item We demonstrate \textbf{completion-throughput gains of up to 2.06$\times$ over DP and 1.7$\times$ over DCP} at matched GPU counts and workloads.
    
\end{itemize}

%% file: Paper/2.Background.tex
\section{Background}

LLM inference consists of two distinct stages: prefill processes the input sequence in parallel, while decode generates subsequent tokens autoregressively. Prefill is typically compute-intensive, whereas long-context decoding places increasing demands on memory bandwidth as attention repeatedly reads the growing key-value (KV) cache~\citep{InfiniteLLM,LongServe,FlashAttention,flashinfer}.

The KV cache avoids recomputing keys and values, but its size grows linearly with batch size, sequence length, layer count, and KV-head count~\citep{vLLM}. Even with cache-reduction techniques~\citep{MQA,GQA}, long contexts and high concurrency place substantial demands on memory capacity and bandwidth.
More broadly, prior GPU-centric systems improve memory efficiency by coordinating data placement and overlapping data movement with computation~\citep{ESPN,espn+}.

\subsection{Inference Optimizations and Disaggregation}

Serving systems improve utilization through iteration-level scheduling~\citep{orca,fastserve,sarathiserve} and paged KV-cache allocation~\citep{vLLM}, while speculative decoding reduces serial generation steps~\citep{Speculativedecoding,Speculativesampling,gammatune,medusa,eagle3}. These optimizations operate within a serving configuration; prefill--decode disaggregation additionally separates the two stages into resource pools that can be provisioned and scaled independently~\citep{distserve,shuffleinfer,beyondbuzz, splitwise}. This separation reduces interference between compute-intensive prefill and iterative decoding, while allowing each pool to use different batching and parallelization strategies. 
TKNP targets execution within the decode pool. The prefill service computes each request's KV cache and transfers it directly to the decode node that will own that request, in TKNP's head layout.

Disaggregation can also separate operators within decoding. Lamina offloads attention and KV cache to memory-oriented devices~\citep{attentionoffloading}, and Hetis distributes attention across heterogeneous GPUs~\citep{hetis}. MegaScale-Infer separates attention and expert FFNs for independent scaling in MoE serving~\citep{megascaleinfer}. KernelFlume targets elastic scaling through expanding attention nodes ~\citep{kernelflume}. 
These designs place weights and attention on separate devices, so weight-holding devices either contribute no KV capacity or have smaller KV budgets than pure attention devices. TKNP instead balances layer-block weights across the same nodes that own request KV cache, so every node contributes approximately equal KV capacity and attention bandwidth, and serves as the dense root for its own blocks.

\subsection{Parallelism Strategies for Inference}

Inference engines combine multiple parallelism primitives to distribute computation and memory. Data parallelism increases throughput through model replicas, but repeated weight storage limits KV-cache capacity per GPU~\citep{pytorchdistributed, vLLM, sglang}. Tensor parallelism shards intra-layer computation and is effective within high-bandwidth interconnect domains~\citep{megatronlm}, while pipeline parallelism distributes consecutive layers across stages and overlaps their execution through microbatching~\citep{gpipe,pipedream}.

When the TP degree exceeds the KV-head count, standard TP replicates KV cache across ranks, limiting useful capacity and resident concurrency. Decode context parallel (DCP) works alongside TP to shard this cache along the sequence dimension instead, combining partial attention results through communication ~\citep{contextparallel,helixparallel,kvparallel,InfiniteLLM,LongServe,ringattention}. 
In current implementations, scaling DCP requires larger TP groups, which extend dense-layer collectives beyond NVLink when they span nodes. In contrast, TKNP retains intra-node TP for projections and MLPs and distributes attention across requests, so each owner computes complete attention outputs for its local head and requests with no cross-node merging of partial results.

Data-parallel attention~\citep{sglang2024dpattention} is closest to TKNP's request-level ownership: each rank computes attention only for its own requests, avoiding KV replication and partial-result merging. However, every rank replicates the attention weights, and the dense layers must span the entire GPU pool, either by sharding weights across nodes with TP, as in Helix~\citep{helixparallel}, which moves dense collectives off NVLink, or by replicating them, which reduces to DP for dense models. TKNP avoids both: only the active root executes each layer's dense operations, within its NVLink domain, while every node computes attention for its own requests. A single, balanced weight copy preserves KV capacity, and cross-node traffic reduces to point-to-point activation transfers sharded by request and head, each about $1/(NT)$ of the hidden state moved by a cross-node TP all-reduce for $N$ nodes of $T$ GPUs (\S\ref{sec:token-parallelism-comms}).

%% file: Paper/3.Motivation.tex
\begin{figure*}[t]
    \centering
    \begin{subfigure}[t]{0.49\textwidth}
        \centering
        \includegraphics[width=\linewidth]{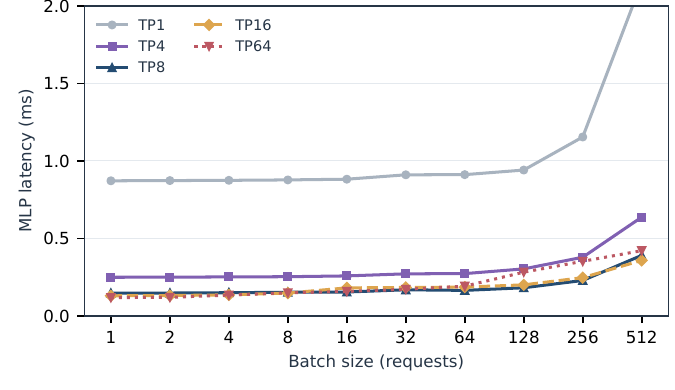}
        \caption{MLP latency, including all-reduce.}
        \label{fig:mlp_tp_decode}
    \end{subfigure}\hfill
    \begin{subfigure}[t]{0.49\textwidth}
        \centering
        \includegraphics[width=\linewidth]{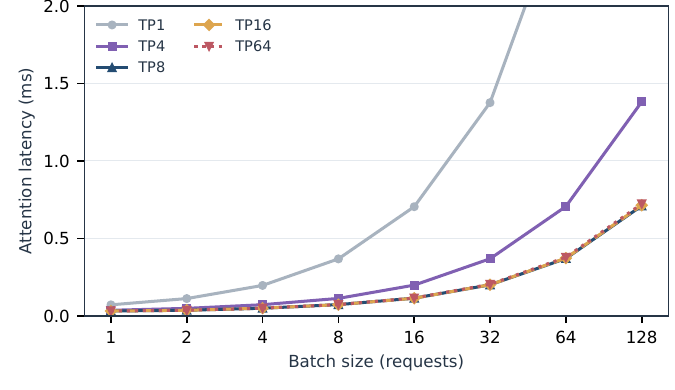}
        \caption{Attention latency with 32K context.}
        \label{fig:attn_tp_decode}
    \end{subfigure}
    \caption{\textbf{Single-layer decode latency for Llama 3.1 405B} on H100 GPUs. (a) TP8 sustains relatively stable MLP latency through batch 128, with larger increases at batches 256--512. (b) Attention grows with batch size; TP8, TP16, and TP64 nearly overlap. Both panels use a 0--2\,ms latency range; TP1 values above this range are clipped.}
    \label{fig:component_batch_scaling}
\end{figure*}

\section{Motivation}

High-throughput decoding requires both efficient execution and enough memory to keep active requests' KV cache resident. 
Using single-layer decode measurements and full-model memory budgets for Llama~3.1~405B on H100 GPUs, we show that dense layers and attention need different forms of parallelism (\S\ref{sec:motiv-scaling}), and that KV capacity, not compute, limits resident concurrency (\S\ref{sec:motiv-memory}).

\subsection{Dense computation and attention need different scaling}
\label{sec:motiv-scaling}

Increasing concurrency changes dense and attention costs differently. 
Figure \ref{fig:component_batch_scaling} shows that at TP8 and 32K cached context, increasing the batch from 1 to 128 raises MLP latency by only 1.24$\times$ but attention latency by 22$\times$.
Attention consequently becomes the dominant component (Figure~\ref{fig:TknpMotivation}). Prior work similarly observes that attention becomes increasingly important as batch size and sequence length grow~\citep{polarsparsity}.
The relatively flat MLP curve reflects the memory-bound regime, in which weight reads dominate latency and are amortized across the batch (Figure~\ref{fig:mlp_tp_decode}).

TP accelerates attention within a node, but gains plateau once its degree reaches the model's eight KV heads. 
At batch 128, attention falls 7.6$\times$ from TP1 to TP8 (5.38 to 0.71\,ms), but an additional 8$\times$ more GPUs at TP64 yields no further improvement (Figure~\ref{fig:attn_tp_decode}).
Replicating KV heads therefore adds GPUs without proportionally increasing attention bandwidth.

\begin{figure}[t]
    \centering
    \includegraphics[width=\linewidth]{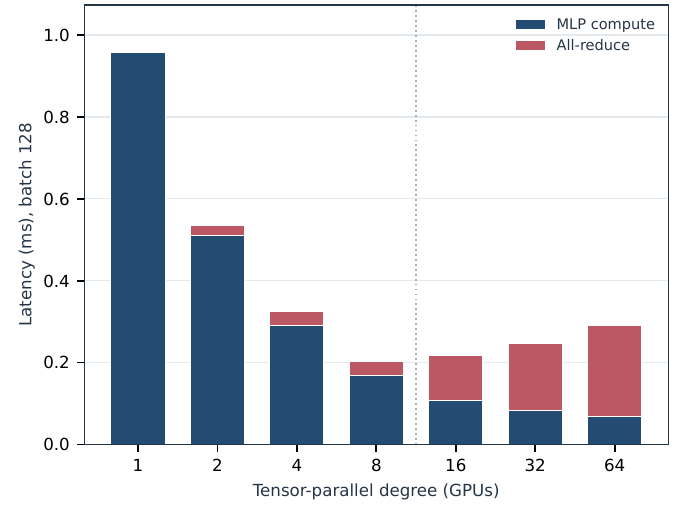}
    \caption{\textbf{Single-layer MLP compute and communication for Llama 3.1 405B} at batch 128. Bars sum local compute measurements and one isolated all-reduce. The dotted line separates single-node and multi-node TP.}
    \label{fig:mlp_compute_collectives}
\end{figure}

Dense computation encounters a different limit when all-reduces cross the inter-node network. 
Figure~\ref{fig:mlp_compute_collectives} shows that at batch 128, one all-reduce costs 6.5$\times$ more at TP64 than at TP8. Despite 8$\times$ less local compute per GPU, end-to-end MLP latency rises by 55\%. DCP inherits this cost, because current implementations expand the TP group across nodes to increase DCP degree. These results favor retaining dense TP within NVLink while expanding attention along another dimension.

PP distributes weights without replication, but does not parallelize attention within a layer. Its microbatches also repeat dense weight reads: splitting batch 128 into four microbatches of 32 invokes each MLP four times. 
Because MLP latency is nearly flat, four batch-32 invocations cost \textbf{3.17$\times$} one batch-128 invocation; pipeline overlap can hide part of this cost but cannot remove the repeated reads.
Efficient decode scaling must therefore account for the different execution and communication requirements of attention and dense operations.

\subsection{Memory efficiency determines feasible concurrency and throughput}
\label{sec:motiv-memory}

\begin{figure}[t]
    \centering
    \includegraphics[width=\linewidth]{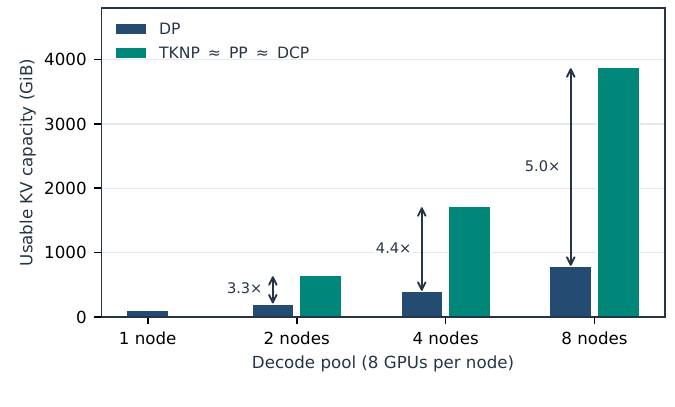}
    \caption{\textbf{Usable KV capacity for Llama 3.1 405B.} Each node contains eight GPUs; DP1 is TP8. TKNP, PP, and DCP provide approximately equal capacity, reaching 5.0$\times$ that of DP on eight nodes.}
    \label{fig:kv_demand_capacity}
\end{figure}

Memory capacity constrains the concurrency that fast decode kernels can sustain. Using FP8 quantization with selected parameters retained in BF16, Llama 3.1 405B model allocations on one eight-GPU TP8 node consume 454\,GiB, leaving only 106\,GiB for KV cache under the measured runtime budget. 
Yet batch 128 at 32K context requires \textbf{2,016\,GiB of BF16 KV cache} before output growth: 19$\times$ a node's KV budget, and 2.4$\times$ the combined budget of eight DP replicas.

As Figure~\ref{fig:kv_demand_capacity} shows, memory-efficient layouts such as PP, DCP, and TKNP provide approximately 5.0$\times$ the usable KV capacity of DP on eight nodes by distributing model weights rather than storing a full model on every node. Standard TP instead spends part of this capacity on replicated KV heads. 

Continuous batching cannot eliminate this capacity constraint: requests that do not fit remain queued, while smaller active batches limit weight-read amortization~\citep{vLLM}. Efficient scaling therefore requires both sufficient KV capacity and an execution strategy that turns it into higher throughput. These observations motivate TKNP's combination of intra-node dense TP, request-parallel attention, and balanced weight placement, allowing the decode pool to expand attention bandwidth while preserving memory for KV cache.

%% file: Paper/4.TokenParallel.tex
\section{Token Parallelism}
\label{sec:token-parallelism}

Decode Token Parallelism separates layer-wise weight placement from request-local attention: dense operations for each layer execute within one node's TP group, while all nodes contribute attention bandwidth and KV capacity at every layer. The base design distributes one model copy across the pool (\S\ref{sec:arch}--\S\ref{sec:token-parallelism-comms}); dual-active-root microbatching adds a second distributed copy to overlap execution (\S\ref{sec:microbatching}).

\subsection{Architecture and Ownership}
\label{sec:arch}

The decode pool contains $N$ nodes of $T$ GPUs, with TP within each NVLink domain and token-parallel communication between matching TP ranks across nodes. An external prefill server places the initial KV cache on the assigned decode owners in the required head layout before generation begins.

In single-copy execution, each node stores a contiguous block of layer weights and becomes the \emph{active root} when that block executes. For layer $\ell$, the active root $b(\ell)$ performs QKV projection, output projection, and MLP computation for the full batch using local TP. Independently, each \emph{request owner} stores its assigned requests' KV cache for every layer and computes their attention, with heads sharded across its TP ranks. Attention therefore spans heads within nodes and requests across nodes, even when the layer's weights reside elsewhere (Figure~\ref{fig:TokenParallel}). This separation is possible because attention over the KV cache is parameter-free: all of a layer's weights reside in its projections and MLP, so a request owner needs only the projected queries, keys, and values for the current token. We call this Token Parallelism because it distributes attention at the granularity of individual decode tokens, each associated with an active request and its locally owned KV cache.

\subsection{Balanced Weight Placement}

\begin{figure*}[t]
    \centering
    \includegraphics[width=\textwidth]{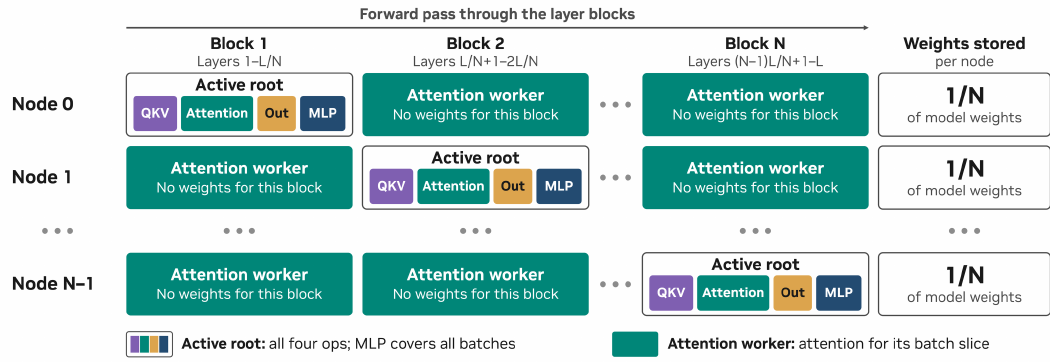}
    \caption{\textbf{Balanced weight placement and changing execution roles.}
    Each node stores one stationary layer block and becomes the active root
    when that block executes. Other nodes compute attention for their assigned
    requests. The schematic shows equal-sized blocks; actual boundaries balance
    weight bytes, including endpoint parameters and quantization differences,
    to leave similar KV-cache budgets.}
    \label{fig:balanced_weight_placement}
\end{figure*}

Separating weights and attention into dedicated pools introduces a capacity tradeoff~\citep{attentionoffloading,kernelflume}: including the weight-owning root in attention creates an uneven KV budget across nodes, while excluding it leaves its spare memory unavailable for KV cache. TKNP resolves this tradeoff by distributing weights across request-owning nodes, allowing the entire pool to contribute approximately equal KV capacity.

\paragraph{Layer-block weight placement.}
TKNP assigns contiguous layer blocks to different nodes and shards each block's weights across local TP ranks (Figure~\ref{fig:balanced_weight_placement}). Each node stores approximately $1/N$ of the model parameters. The block's node becomes the active root and executes projections and MLPs for the full batch, while every node computes attention for its assigned requests. This preserves full-batch weight-read amortization. The active-root role follows the blocks, while weights and KV cache remain stationary.

\paragraph{Capacity-aware request assignment.}
The scheduler assigns each waiting request to an eligible node with the lowest KV-cache load, measured by current usage or reserved context length. Eligibility requires sufficient capacity across the node’s TP ranks for the input and expected output growth. If no node has enough capacity, the request remains queued. Once assigned, ownership stays fixed until completion, avoiding KV-cache migration. Because decode attention cost scales with resident KV length, balancing KV load also balances per-layer attention time.

\subsection{Decode Execution and Communication}

At each layer, the active root computes QKV projections for the full batch using local TP, then scatters the projected tensors to their request owners (Figure~\ref{fig:TokenParallel}). All nodes, including root, apply rotary embeddings, append the new keys and values to local KV cache, and compute attention for their assigned requests and heads. The root gathers the attention outputs in batch order and completes the output projection, residual and normalization operations, and MLP. Each owner holds the complete KV cache for its heads, so attention requires no cross-node reduction of partial results, while dense all-reduces remain within NVLink.

Execution stays on the active root throughout its layer block. At each block boundary, hidden and residual activations pass to the next weight-owning node, which becomes the new root. 

\subsection{Token Parallel Communication Efficiency}
\label{sec:token-parallelism-comms}

TKNP confines cross-node attention exchange to peer-wise scatter/gather between matching TP ranks on the active root and request owners (Figure~\ref{fig:TokenParallel}). Dense all-reduces stay within NVLink, and request-owned attention needs no cross-node aggregation of partial softmax results. This separates attention scaling from the cross-node collectives incurred when DCP expands the dense TP domain.

Let $B$ be the active batch size, $H_q$ and $H_{kv}$ the query and KV-head counts, and $d_h$ the head dimension. With intra-node TP degree $T$, each root rank holds projected QKV of shape
\begin{equation}
X^{\mathrm{qkv}}\in\mathbb{R}^{B\times (H_q+2H_{kv})d_h/T}.
\end{equation}
An owner assigned $B_g$ requests receives and returns tensors of shapes
\begin{equation}
\begin{aligned}
\text{scatter:}\quad &[B_g,(H_q+2H_{kv})d_h/T],\\
\text{gather:}\quad &[B_g,H_q/T,d_h].
\end{aligned}
\end{equation}
Balanced assignment gives $B_g\approx B/N$; capacity-aware scheduling determines the actual split. The root's own requests require no network transfer. For $s$ bytes per element, the combined scatter/gather payload per layer and matching-rank group is
\begin{equation}
C_\ell=\frac{2s(H_q+H_{kv})d_h}{T}
\sum_{g\ne b(\ell)}B_g.
\end{equation}
This counts each payload once, excluding padding, protocol overhead, and block handoffs. Individual transfers are sharded in both batch and head dimensions, and their volume is independent of context length.

For DCP degree $N$ combined with TP degree $NT$, each rank computes partial attention over a sequence shard for the full batch. Its aggregation tensors have shapes
\begin{equation}
\begin{aligned}
O_{\mathrm{DCP}}&\in\mathbb{R}^{B\times H_q/T\times d_h},\\
L_{\mathrm{DCP}}&\in\mathbb{R}^{B\times H_q/T},
\end{aligned}
\end{equation}
where $L_{\mathrm{DCP}}$ contains FP32 log-sum-exp statistics. DCP all-gathers these statistics and reduce-scatters the corrected partial outputs, whereas TKNP gathers complete outputs for only $B_g$ requests. Under balanced assignment, the output tensor-size ratio is
\begin{equation}
\frac{|O_{\mathrm{TKNP},g}|}{|O_{\mathrm{DCP}}|}
=\frac{B_g}{B}\approx\frac{1}{N}.
\end{equation}
For Llama 3.1 405B at batch 128 on four TP8 nodes, BF16 partial outputs occupy 512\,KiB per DCP rank, compared with 128\,KiB per remote TKNP owner, a 75\% reduction in output tensor size. DCP additionally exchanges 8\,KiB of LSE statistics per rank; TKNP instead dispatches 144\,KiB of QKV per remote owner. The $N$-fold reduction concerns individual output tensors, not total network traffic: DCP also gathers queries, and collective algorithms determine transfer volume.

By comparison, dense TP all-reduces operate on full-batch hidden activations of shape $[B,D]$, where $D$ is the hidden dimension. Since $H_q d_h = D$ for both evaluated models, each TKNP gather is $1/(NT)$ of this operand: 128\,KiB versus 4\,MiB on four TP8 nodes.
When TP spans nodes alongside DCP, these collectives traverse the inter-node network in addition to communication for sequence-sharded attention. TKNP instead keeps dense collectives local and exchanges projected head-space tensors across nodes, preserving the dense scaling benefits measured in Figure~\ref{fig:mlp_compute_collectives}. 

\subsection{Dual-Active-Root Microbatching}
\label{sec:microbatching}

\begin{figure}[t]
    \centering
    \includegraphics[width=\linewidth]{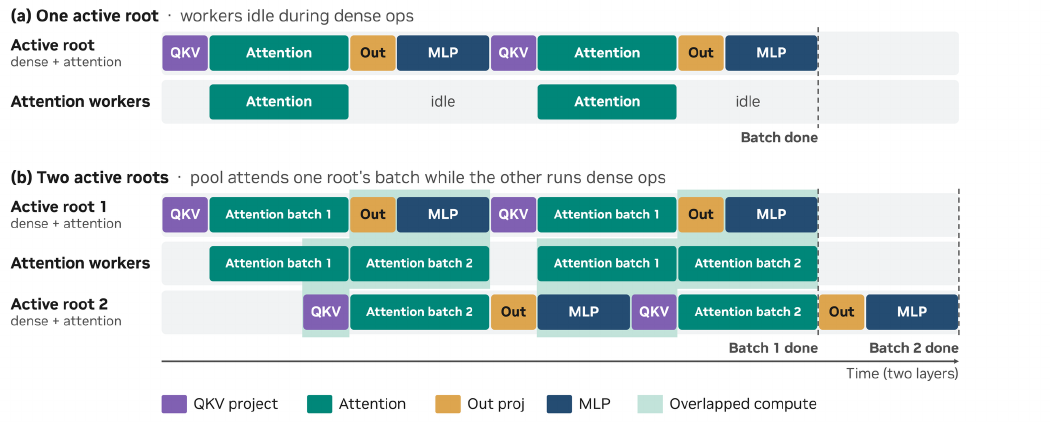}
    \caption{\textbf{Dual-active-root microbatching.}
    Shared workers alternate attention between two microbatches to overlap it with dense computation on the other root.}
    \label{fig:two_active_roots_microbatch}
\end{figure}

Although attention dominates long-context decoding, attention workers still encounter brief idle gaps while the active root executes projections and the MLP. In larger decode pools, such as those with eight or more nodes, the additional KV capacity can make partial weight replication a favorable tradeoff, enabling greater compute overlap at the cost of some KV-cache capacity. For this regime, we propose splitting the batch into two independent microbatches served by distinct active roots. As shown in Figure~\ref{fig:two_active_roots_microbatch}, shared workers alternate attention between the two microbatches, using one microbatch's attention to fill gaps during the other's dense computation. This overlap offers an opportunity to improve throughput when the additional weight storage does not constrain request admission.

For even $N$, block $b$ is stored on nodes $b$ and $(b+N/2)\bmod N$, providing two distinct roots for each block. Every node stores two blocks, approximately $2/N$ of a model's weights instead of $1/N$, with endpoint parameters included in the balancing budget. This shifted placement distributes the second copy across the pool while keeping weights stationary as root roles rotate.
Each request belongs to one microbatch and keeps a single owner for all layers. While a node serves as one microbatch's active root, it computes attention only for that microbatch's requests, so the two roots never contend for dense and attention work in the same phase. KV cache is neither migrated nor replicated.

%% file: Paper/5.Evaluation.tex
\section{Evaluation}
\label{sec:evaluation}

We integrate TKNP on top of vLLM by extending model construction and weight loading to instantiate dense parameters only on their assigned nodes, while retaining attention and local KV cache on every node. The scheduler tracks request ownership and per-owner KV budgets; the model runner maps scheduled requests into owner-specific execution slots and prepares local attention metadata. Dedicated communication groups connect matching TP ranks across nodes for scatter/gather and block handoffs, while reusing existing attention kernels and intra-node TP. Full CUDA-graph replay covers this distributed decode path through captured ownership layouts. 

We evaluate how TKNP's combination of KV capacity and distributed attention affects decode throughput and request latency. The central question is when memory-efficient concurrency outweighs the execution advantage of independent replicas.

\begin{figure*}[t]
    \centering
    \begin{subfigure}[t]{0.325\textwidth}
        \includegraphics[width=\linewidth]{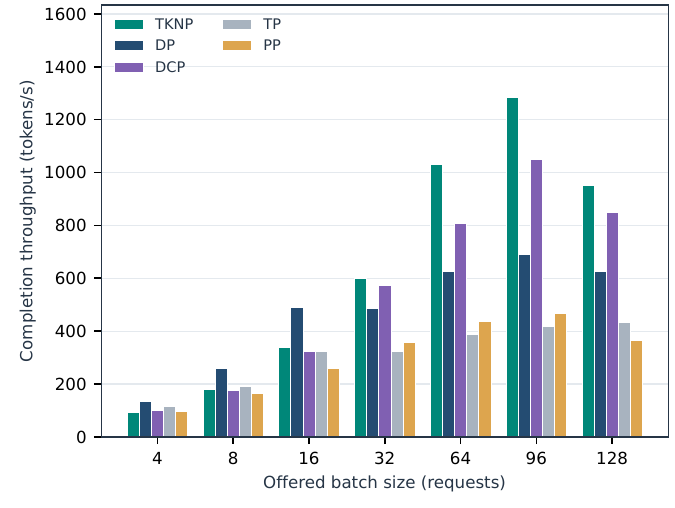}
        \caption{32K context.}
    \end{subfigure}\hfill
    \begin{subfigure}[t]{0.325\textwidth}
        \includegraphics[width=\linewidth]{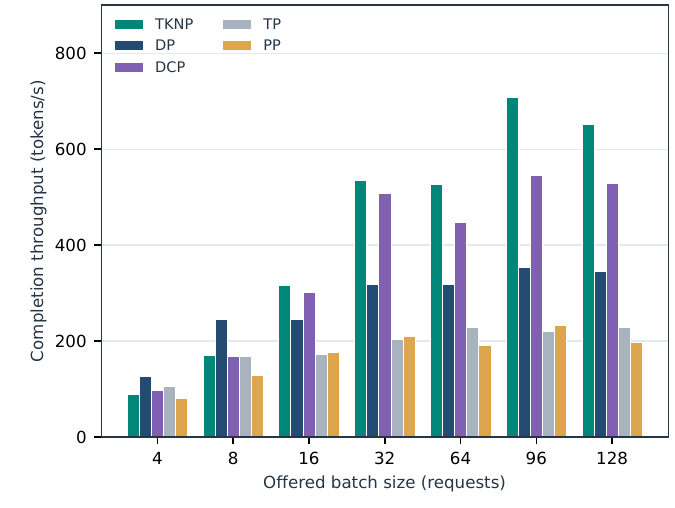}
        \caption{64K context.}
    \end{subfigure}\hfill
    \begin{subfigure}[t]{0.325\textwidth}
        \includegraphics[width=\linewidth]{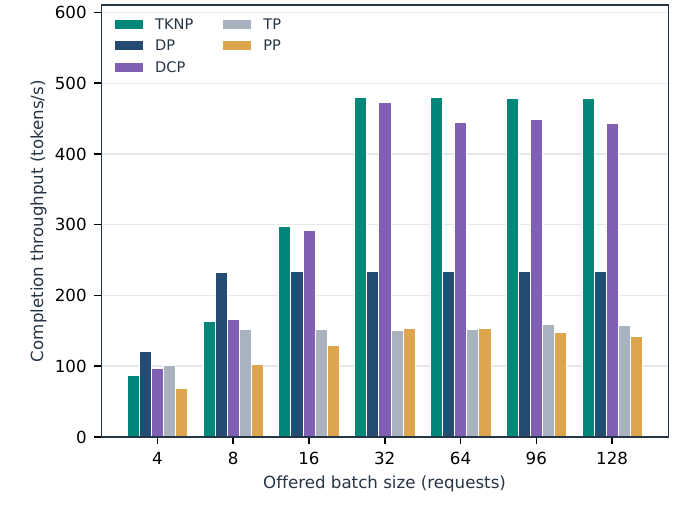}
        \caption{96K context.}
    \end{subfigure}
    \caption{\textbf{Llama 3.1 405B completion throughput on 32 H100 GPUs}, including admission, capacity-induced waves, and drain, with 1,024 output tokens per request. While DP leads at low concurrency, TKNP achieves up to 2.06$\times$ speedups on large batch, long sequence workloads.}
    \label{fig:eval_llama_throughput}
    \par\medskip
    \setcounter{subfigure}{0}
    \centering
    \begin{subfigure}[t]{0.325\textwidth}
        \includegraphics[width=\linewidth]{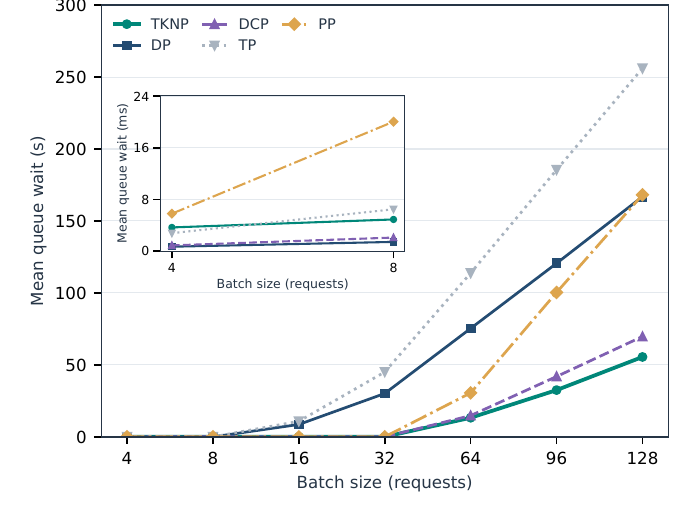}
        \caption{Mean queue wait.}
    \end{subfigure}\hfill
    \begin{subfigure}[t]{0.325\textwidth}
        \includegraphics[width=\linewidth]{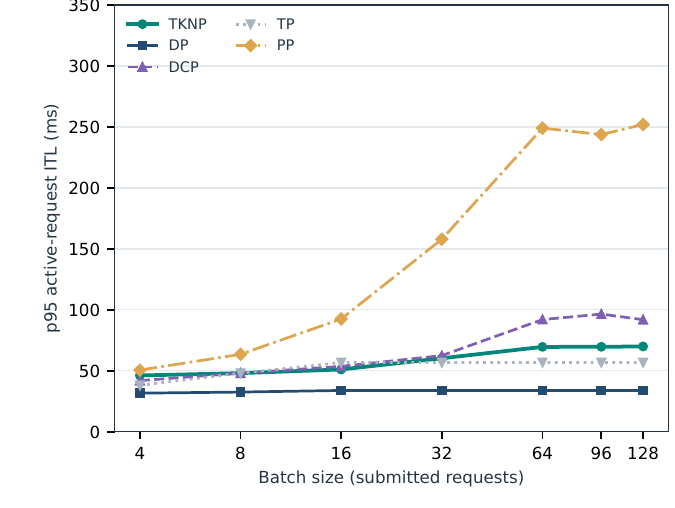}
        \caption{p95 active-request ITL.}
        \label{fig:eval_active_itl}
    \end{subfigure}\hfill
    \begin{subfigure}[t]{0.325\textwidth}
        \includegraphics[width=\linewidth]{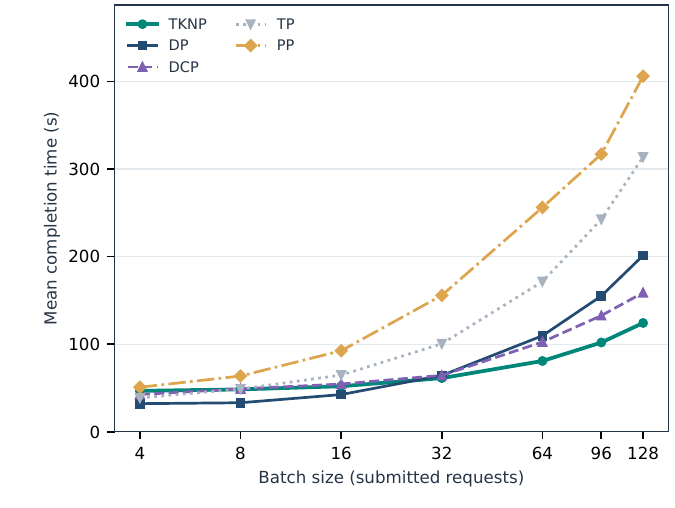}
        \caption{Mean request completion time.}
        \label{fig:eval_mean_completion}
    \end{subfigure}
    \caption{\textbf{Admission waiting, active-request ITL, and average completion latency} for Llama 3.1 405B at 64K context on 32 H100 GPUs, with 1K OSL. The x-axis counts all submitted requests; only a capacity-limited subset may be active. (a) Release to first scheduling. (b) Token intervals after the first token, excluding initial queue wait. (c) Average request completion time including queueing.}
    \label{fig:eval_latency_tradeoff}
\end{figure*}

\begin{figure}[t]
    \centering
    \includegraphics[width=0.80\linewidth]{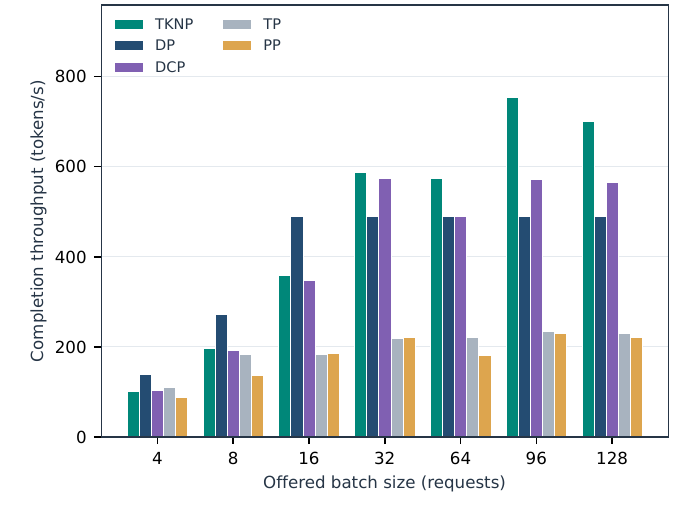}
    \caption{\textbf{Nemotron-4 340B completion throughput} at 64K context on 32 H100 GPUs, with 1,024 output tokens per request, including all admission waves.}
    \label{fig:eval_nemotron_throughput}
\end{figure}

\begin{figure}[t]
    \centering
    \includegraphics[width=0.80\linewidth]{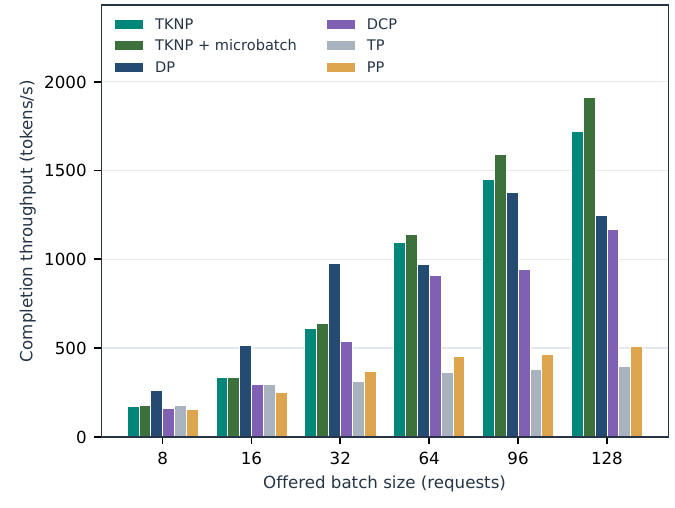}
    \caption{\textbf{Eight-node decode throughput with microbatching} for Llama 3.1 405B at 32K context on 64 H100 GPUs, with 1,024 output tokens per request. TKNP denotes the single-copy schedule; TKNP + microbatch uses two active roots.}
    \label{fig:eval_llama_8node_throughput}
\end{figure}

\subsection{Experimental Setup}

\paragraph{Disaggregated serving configuration.}

We consider a prefill–decode disaggregated setup with two prefill nodes and a decode pool of four or eight nodes. Each node contains 8$\times$80GB H100 GPUs connected via 4th-generation NVLink and dual Intel Sapphire Rapids CPUs, with nodes interconnected by 8$\times$400 Gb/s InfiniBand.  We evaluate Llama 3.1 405B and Nemotron-4 340B using FP8 quantization with selected parameters retained in BF16, BF16 KV cache, and CUDA graphs on top of vLLM. Nemotron uses linear RoPE scaling beyond its original context window; these results characterize decode performance rather than long-context accuracy. Our software stack includes PyTorch 2.9.1, CUDA 12.9, NCCL 2.27.5, vLLM 0.19.0, and FlashAttention-3 \cite{FlashAttention3}, running on Linux kernel 5.15.0-88.

\paragraph{Parallel configurations.}
All experiments use the same two-node prefill configuration with TP8$\times$DP2, isolating performance differences to the decode configuration. For $N$ decode nodes, TKNP uses TP8 within each node and distributes weights and request-owned KV across the pool. DP uses $N$ independent TP8 replicas; TP spans all $8N$ GPUs; DCP combines TP$8N$ with DCP$N$, since vLLM's DCP reuses the TP group; and PP uses TP8 within each of $N$ stages. PP divides the resident batch into approximately $N$ microbatches. All comparisons use matched decode GPU counts and precision. Completion throughput and latency are measured at the decode pool from workload release; each request generates exactly the specified output length; and under greedy decoding, TKNP's generated tokens match those of the DP baseline.

\paragraph{Workload construction.}
\label{sec:workloads}

We evaluate our system using two workload pools. First, we generate synthetic long-context workloads using RULER ~\citep{ruler} with controlled input (ISL) and output sequence lengths (OSL) to study performance across varying concurrency and context lengths. For each context length, we hold the sequence length fixed while varying batch size, with each batch comprising distinct, non-identical requests. We disable prefix sharing across requests throughout our evaluation.

Second, we construct a realistic mixed-length workload from L-Eval, LongBench v1, and LongBench v2 ~\citep{leval,longbench,longbenchv2}, with sequence lengths ranging from 8,223 to 126,848 tokens (mean 32,898, median 19,048, standard deviation 28,496). 
The workload contains substantial variation in sequence length: 12.5\% of requests exceed 64K tokens and contribute 36.1\% of all input tokens, while the longest 25\% account for 57.2\%. This distribution captures heterogeneous KV-cache demand and attention computation across requests. Following prior work, we evaluate online serving under Poisson arrivals at varying request rates and a synchronized burst where all requests are submitted simultaneously \citep{distserve}. All systems use the same generated arrival trace at each request rate.

\subsection{Performance across Concurrency and Context Length}
\label{sec:burst-workloads}

We evaluate performance across varying concurrency levels and context lengths using fixed-length workloads, with each request generating 1,024 output tokens. We define completion throughput as the total number of generated output tokens divided by the elapsed time from workload release until the final request completes, including admission waiting, all execution waves, and final drain.

DP performs best when the submitted workload fits within per-replica KV-cache capacity. In this regime, replicas execute independently without TKNP’s per-layer scatter/gather overhead, while small active batches provide limited attention work to parallelize. As concurrency increases, however, replicated model weights constrain DP’s KV-cache capacity and reduce the number of requests that can execute concurrently. Excess requests are deferred to subsequent admission waves, limiting effective batch size and overall throughput. TKNP instead uses the distributed memory capacity to retain larger active batches while parallelizing attention across the decode pool. Consequently, the crossover from DP to TKNP occurs at lower concurrency as context length increases and KV capacity becomes constrained earlier (Figure~\ref{fig:eval_llama_throughput}). TKNP improves completion throughput by up to 2.06$\times$ over DP.

Comparisons with the remaining baselines further separate the effects of KV capacity and execution efficiency. TP replicates KV heads while introducing cross-node collectives for dense layers, limiting both capacity and execution efficiency. PP preserves aggregate KV capacity but partitions dense computation into smaller stage-local microbatches and confines each layer’s attention computation to a single pipeline stage. DCP provides the most direct capacity-matched comparison: like TKNP, it shards the KV cache across the decode pool, but requires partial-attention aggregation and additional inter-node communication. TKNP therefore achieves its advantage over DCP with comparable KV capacity, indicating that the gains arise from more efficient distributed execution rather than additional cache capacity. On eight nodes, single-copy TKNP improves completion throughput by up to 1.53$\times$ over DCP; dual-active-root microbatching raises this advantage to 1.7$\times$ (Figure~\ref{fig:eval_llama_8node_throughput}).

Nemotron-4 340B's smaller weight footprint leaves more KV capacity, improving DP's resident concurrency and competitiveness. TKNP nevertheless retains the high-concurrency advantage, achieving up to 1.54$\times$ higher throughput than DP and 1.32$\times$ than DCP (Figure~\ref{fig:eval_nemotron_throughput}).

Figure~\ref{fig:eval_latency_tradeoff} separates the latency experienced by admitted requests from the end-to-end latency experienced by all submitted requests. Although the x-axis represents the full submitted batch, KV-cache constraints allow only a subset of requests to decode concurrently. Consequently, the lower active-request ITL of DP and TP in Figure~\ref{fig:eval_active_itl} partly reflects their smaller resident batches and excludes the admission delay experienced by queued requests. Mean completion time (Figure~\ref{fig:eval_mean_completion}) captures both effects by measuring from request release through completion. At high concurrency, TKNP’s greater resident capacity substantially reduces queuing and completion time even when DP achieves lower ITL for requests already admitted. Capacity alone, however, does not guarantee lower latency: PP admits larger batches than DP but can exhibit similar queueing delay because each wave takes longer to complete. TKNP combines greater resident capacity with efficient distributed execution for improving end-to-end request completion.

\begin{figure*}[!t]
    \centering
    \begin{subfigure}[t]{0.325\textwidth}
        \includegraphics[width=\linewidth]{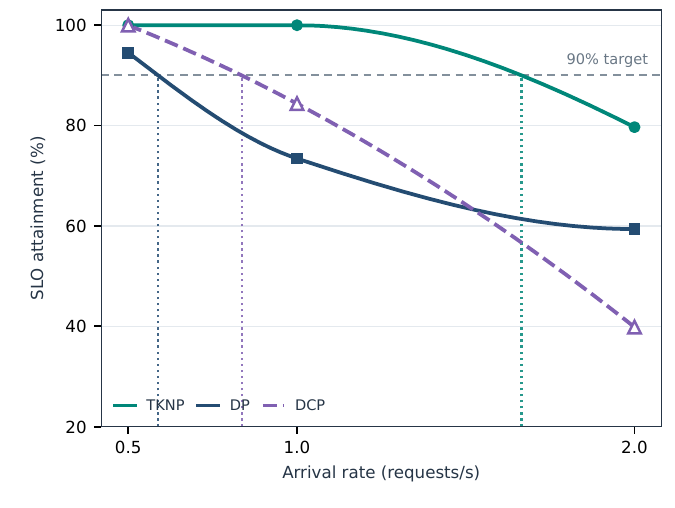}
        \caption{SLO attainment.}
    \end{subfigure}\hfill
    \begin{subfigure}[t]{0.325\textwidth}
        \includegraphics[width=\linewidth]{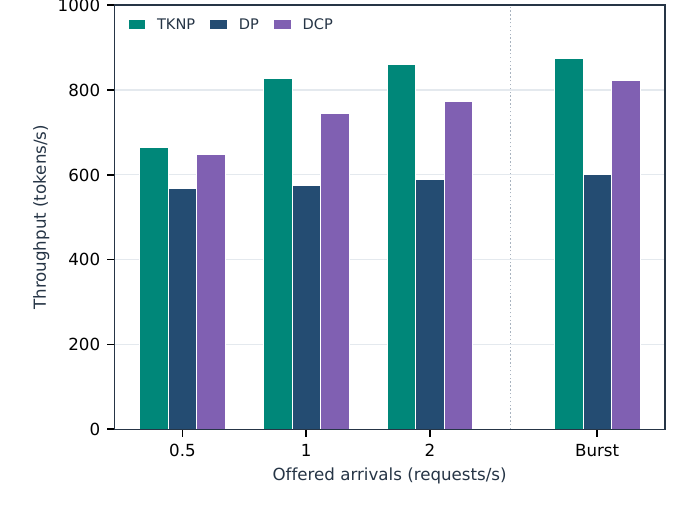}
        \caption{Throughput.}
    \end{subfigure}\hfill
    \begin{subfigure}[t]{0.325\textwidth}
        \includegraphics[width=\linewidth]{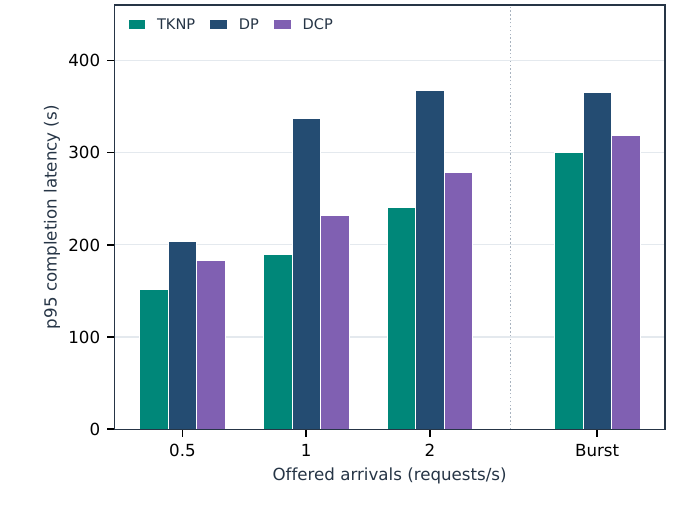}
        \caption{p95 request completion time.}
    \end{subfigure}
    \caption{\textbf{TKNP sustains its throughput advantage under mixed input lengths.} Llama 3.1 405B on 32 H100 GPUs. (a) SLO: arrival-to-completion time per output token $\leq$100 ms. (b,c) Completion throughput and p95 request completion time under Poisson arrivals and synchronized bursts.}
    \label{fig:eval_mixed_workload}
\end{figure*}

\subsection{Scaling and Microbatching}
\label{sec:microbatch-evaluation}

For the same 128-request, 32K workload, doubling the decode pool from four to eight nodes increases single-copy TKNP throughput by 1.81$\times$, corresponding to 90.5\% scaling efficiency, compared with 1.37$\times$ for DCP (Figures~\ref{fig:eval_llama_throughput} and~\ref{fig:eval_llama_8node_throughput}). These gains include reduced admission waves as well as faster execution. DP nearly doubles throughput as its aggregate KV capacity grows, becoming more competitive with DCP, whose attention aggregation and dense collectives span the larger inter-node group. By keeping dense collectives within each node and using inter-node scatter/gather for attention, TKNP retains higher throughput at large batches and scales more effectively than DCP.

Microbatching improves throughput by up to 11.1\%, reaching 1.7$\times$ the throughput of DCP, with the largest gains at high concurrency. The overlap opportunity is limited to the brief gaps between attention phases, since attention already dominates execution. Additional microbatch scheduling and synchronization, smaller dense batches with less weight-read amortization, and reduced KV capacity from the second weight copy can further offset the benefit. These tradeoffs favor single-copy TKNP in smaller pools where memory limits resident concurrency.

\subsection{Mixed-Length Serving Workloads}
\label{sec:mixed-workload}

We next evaluate whether TKNP maintains its efficiency under the heterogeneous mixed-length workload described in Section~\ref{sec:workloads}.
Despite substantial variation in per-request KV demand and attention work, 
TKNP achieves the highest throughput at every arrival rate, preserving the trends observed with fixed-length workloads (Figure~\ref{fig:eval_mixed_workload}b). This demonstrates that TKNP remains effective under realistic variation in request lengths.

The same behavior persists under staggered arrivals, where TKNP sustains high throughput while maintaining strong tail completion latency (Figure~\ref{fig:eval_mixed_workload}c). 
We define the completion SLO as arrival-to-completion latency normalized by the number of generated output tokens, including queueing time. At a 100 ms/output-token SLO, TKNP achieves full attainment at one request/s, while DP and DCP fall below 90\% (Figure~\ref{fig:eval_mixed_workload}a). At two requests/s, TKNP retains higher SLO attainment than both baselines alongside its aggregate throughput advantage. Overall, TKNP maintains its efficiency across both synchronized and staggered arrivals, showing that its benefits generalize to heterogeneous serving workloads.

%% file: Paper/6.FutureWork.tex
\section{Limitations and Future Work}

TKNP targets dense-model serving under concurrent, long-context decoding. At small batches or short contexts, communication and synchronization can outweigh the benefit of distributed attention. For the targeted workloads, intra-node TP provides effective scaling for projections and MLP computation, while TKNP extends attention bandwidth and KV capacity across nodes. Finally, our evaluation uses H100 nodes connected by InfiniBand; larger NVLink domains shift the balance between intra- and inter-node communication.

Several directions could extend TKNP. Applying TKNP's balanced layer-block placement to DCP would keep dense collectives within each node while sharding the sequence, removing DCP's need for cross-node TP and supporting requests whose KV cache exceeds one node. Combining TKNP with expert parallelism could reduce the attention-weight replication of data-parallel attention in mixture-of-experts serving. Dual-active-root microbatching improves throughput by up to 11.1\% but adds scheduling, synchronization, and placement complexity. Adaptive schedulers that jointly account for KV occupancy and attention cost could capture this overlap with less memory overhead.

%% file: Paper/7.Conclusion.tex
\section{Conclusion}

Decode Token Parallelism builds on the observation that dense layers and attention scale differently during decode. It decouples weight placement from attention ownership: each layer's dense operations run on a single node within its NVLink domain, while every node computes request-local attention against its own KV cache, and a single, balanced weight copy preserves KV capacity across the pool. On Llama~3.1~405B and Nemotron-4~340B, TKNP improves completion throughput by up to 1.7$\times$ over DCP and by up to 2.06$\times$ over DP. Dual-active-root microbatching further improves base TKNP throughput by up to 11.1\% at larger decode pools through compute–attention overlap. Overall, TKNP demonstrates that decoupling attention scaling from model-weight placement enables both memory-efficient and high-throughput long-context decoding at scale.

%% file: references.bib
@inproceedings {orca,
author = {Gyeong-In Yu and Joo Seong Jeong and Geon-Woo Kim and Soojeong Kim and Byung-Gon Chun},
title = {Orca: A Distributed Serving System for {Transformer-Based} Generative Models},
booktitle = {16th USENIX Symposium on Operating Systems Design and Implementation (OSDI 22)},
year = {2022},
isbn = {978-1-939133-28-1},
address = {Carlsbad, CA},
pages = {521--538},
url = {https://www.usenix.org/conference/osdi22/presentation/yu},
publisher = {USENIX Association},
month = jul
}

@inproceedings{vLLM,
author = {Kwon, Woosuk and Li, Zhuohan and Zhuang, Siyuan and Sheng, Ying and Zheng, Lianmin and Yu, Cody Hao and Gonzalez, Joseph and Zhang, Hao and Stoica, Ion},
title = {Efficient Memory Management for Large Language Model Serving with PagedAttention},
year = {2023},
isbn = {9798400702297},
publisher = {Association for Computing Machinery},
address = {New York, NY, USA},
url = {https://doi.org/10.1145/3600006.3613165},
doi = {10.1145/3600006.3613165},
booktitle = {Proceedings of the 29th Symposium on Operating Systems Principles},
pages = {611–626},
numpages = {16},
location = {Koblenz, Germany},
series = {SOSP '23}
}

@misc{fastserve,
      title={Fast Distributed Inference Serving for Large Language Models}, 
      author={Bingyang Wu and Yinmin Zhong and Zili Zhang and Shengyu Liu and Fangyue Liu and Yuanhang Sun and Gang Huang and Xuanzhe Liu and Xin Jin},
      year={2024},
      eprint={2305.05920},
      archivePrefix={arXiv},
      primaryClass={cs.LG},
      url={https://arxiv.org/abs/2305.05920}, 
}

@inproceedings{distserve,
author = {Zhong, Yinmin and Liu, Shengyu and Chen, Junda and Hu, Jianbo and Zhu, Yibo and Liu, Xuanzhe and Jin, Xin and Zhang, Hao},
title = {DistServe: disaggregating prefill and decoding for goodput-optimized large language model serving},
year = {2024},
isbn = {978-1-939133-40-3},
publisher = {USENIX Association},
address = {USA},
booktitle = {Proceedings of the 18th USENIX Conference on Operating Systems Design and Implementation},
articleno = {11},
numpages = {18},
location = {Santa Clara, CA, USA},
series = {OSDI'24}
}

@misc{attentionoffloading,
      title={Efficient Heterogeneous Large Language Model Decoding with Model-Attention Disaggregation}, 
      author={Shaoyuan Chen and Wencong Xiao and Yutong Lin and Mingxing Zhang and Yingdi Shan and Jinlei Jiang and Kang Chen and Yongwei Wu},
      year={2025},
      eprint={2405.01814},
      archivePrefix={arXiv},
      primaryClass={cs.LG},
      url={https://arxiv.org/abs/2405.01814}, 
}

@inproceedings{megascaleinfer,
author = {Zhu, Ruidong and Jiang, Ziheng and Jin, Chao and Wu, Peng and Stuardo, Cesar A. and Wang, Dongyang and Zhang, Xinlei and Zhou, Huaping and Wei, Haoran and Cheng, Yang and Xiao, Jianzhe and Zhang, Xinyi and Liu, Lingjun and Lin, Haibin and Chang, Li-Wen and Ye, Jianxi and Yu, Xiao and Liu, Xuanzhe and Jin, Xin and Liu, Xin},
title = {MegaScale-Infer: Efficient Mixture-of-Experts Model Serving with Disaggregated Expert Parallelism},
year = {2025},
isbn = {9798400715242},
publisher = {Association for Computing Machinery},
address = {New York, NY, USA},
url = {https://doi.org/10.1145/3718958.3750506},
doi = {10.1145/3718958.3750506},
booktitle = {Proceedings of the ACM SIGCOMM 2025 Conference},
pages = {592–608},
numpages = {17},
location = {S\~{a}o Francisco Convent, Coimbra, Portugal},
series = {SIGCOMM '25}
}

@misc{contextparallel,
      title={Context Parallelism for Scalable Million-Token Inference}, 
      author={Amy Yang and Jingyi Yang and Aya Ibrahim and Xinfeng Xie and Bangsheng Tang and Grigory Sizov and Jeremy Reizenstein and Jongsoo Park and Jianyu Huang},
      year={2025},
      eprint={2411.01783},
      archivePrefix={arXiv},
      primaryClass={cs.DC},
      url={https://arxiv.org/abs/2411.01783}, 
}

@misc{llama3,
      title={The Llama 3 Herd of Models}, 
      author={Aaron Grattafiori and Abhimanyu Dubey and Abhinav Jauhri and Abhinav Pandey and Abhishek Kadian and Ahmad Al-Dahle and Aiesha Letman and Akhil Mathur and Alan Schelten and Alex Vaughan and Amy Yang and Angela Fan and Anirudh Goyal and Anthony Hartshorn and Aobo Yang and Archi Mitra and Archie Sravankumar and Artem Korenev and Arthur Hinsvark and Arun Rao and Aston Zhang and Aurelien Rodriguez and Austen Gregerson and Ava Spataru and Baptiste Roziere and Bethany Biron and Binh Tang and Bobbie Chern and Charlotte Caucheteux and Chaya Nayak and Chloe Bi and Chris Marra and Chris McConnell and Christian Keller and Christophe Touret and Chunyang Wu and Corinne Wong and Cristian Canton Ferrer and Cyrus Nikolaidis and Damien Allonsius and Daniel Song and Danielle Pintz and Danny Livshits and Danny Wyatt and David Esiobu and Dhruv Choudhary and Dhruv Mahajan and Diego Garcia-Olano and Diego Perino and Dieuwke Hupkes and Egor Lakomkin and Ehab AlBadawy and Elina Lobanova and Emily Dinan and Eric Michael Smith and Filip Radenovic and Francisco Guzmán and Frank Zhang and Gabriel Synnaeve and Gabrielle Lee and Georgia Lewis Anderson and Govind Thattai and Graeme Nail and Gregoire Mialon and Guan Pang and Guillem Cucurell and Hailey Nguyen and Hannah Korevaar and Hu Xu and Hugo Touvron and Iliyan Zarov and Imanol Arrieta Ibarra and Isabel Kloumann and Ishan Misra and Ivan Evtimov and Jack Zhang and Jade Copet and Jaewon Lee and Jan Geffert and Jana Vranes and Jason Park and Jay Mahadeokar and Jeet Shah and Jelmer van der Linde and Jennifer Billock and Jenny Hong and Jenya Lee and Jeremy Fu and Jianfeng Chi and Jianyu Huang and Jiawen Liu and Jie Wang and Jiecao Yu and Joanna Bitton and Joe Spisak and Jongsoo Park and Joseph Rocca and Joshua Johnstun and Joshua Saxe and Junteng Jia and Kalyan Vasuden Alwala and Karthik Prasad and Kartikeya Upasani and Kate Plawiak and Ke Li and Kenneth Heafield and Kevin Stone and Khalid El-Arini and Krithika Iyer and Kshitiz Malik and Kuenley Chiu and Kunal Bhalla and Kushal Lakhotia and Lauren Rantala-Yeary and Laurens van der Maaten and Lawrence Chen and Liang Tan and Liz Jenkins and Louis Martin and Lovish Madaan and Lubo Malo and Lukas Blecher and Lukas Landzaat and Luke de Oliveira and Madeline Muzzi and Mahesh Pasupuleti and Mannat Singh and Manohar Paluri and Marcin Kardas and Maria Tsimpoukelli and Mathew Oldham and Mathieu Rita and Maya Pavlova and Melanie Kambadur and Mike Lewis and Min Si and Mitesh Kumar Singh and Mona Hassan and Naman Goyal and Narjes Torabi and Nikolay Bashlykov and Nikolay Bogoychev and Niladri Chatterji and Ning Zhang and Olivier Duchenne and Onur Çelebi and Patrick Alrassy and Pengchuan Zhang and Pengwei Li and Petar Vasic and Peter Weng and Prajjwal Bhargava and Pratik Dubal and Praveen Krishnan and Punit Singh Koura and Puxin Xu and Qing He and Qingxiao Dong and Ragavan Srinivasan and Raj Ganapathy and Ramon Calderer and Ricardo Silveira Cabral and Robert Stojnic and Roberta Raileanu and Rohan Maheswari and Rohit Girdhar and Rohit Patel and Romain Sauvestre and Ronnie Polidoro and Roshan Sumbaly and Ross Taylor and Ruan Silva and Rui Hou and Rui Wang and Saghar Hosseini and Sahana Chennabasappa and Sanjay Singh and Sean Bell and Seohyun Sonia Kim and Sergey Edunov and Shaoliang Nie and Sharan Narang and Sharath Raparthy and Sheng Shen and Shengye Wan and Shruti Bhosale and Shun Zhang and Simon Vandenhende and Soumya Batra and Spencer Whitman and Sten Sootla and Stephane Collot and Suchin Gururangan and Sydney Borodinsky and Tamar Herman and Tara Fowler and Tarek Sheasha and Thomas Georgiou and Thomas Scialom and Tobias Speckbacher and Todor Mihaylov and Tong Xiao and Ujjwal Karn and Vedanuj Goswami and Vibhor Gupta and Vignesh Ramanathan and Viktor Kerkez and Vincent Gonguet and Virginie Do and Vish Vogeti and Vítor Albiero and Vladan Petrovic and Weiwei Chu and Wenhan Xiong and Wenyin Fu and Whitney Meers and Xavier Martinet and Xiaodong Wang and Xiaofang Wang and Xiaoqing Ellen Tan and Xide Xia and Xinfeng Xie and Xuchao Jia and Xuewei Wang and Yaelle Goldschlag and Yashesh Gaur and Yasmine Babaei and Yi Wen and Yiwen Song and Yuchen Zhang and Yue Li and Yuning Mao and Zacharie Delpierre Coudert and Zheng Yan and Zhengxing Chen and Zoe Papakipos and Aaditya Singh and Aayushi Srivastava and Abha Jain and Adam Kelsey and Adam Shajnfeld and Adithya Gangidi and Adolfo Victoria and Ahuva Goldstand and Ajay Menon and Ajay Sharma and Alex Boesenberg and Alexei Baevski and Allie Feinstein and Amanda Kallet and Amit Sangani and Amos Teo and Anam Yunus and Andrei Lupu and Andres Alvarado and Andrew Caples and Andrew Gu and Andrew Ho and Andrew Poulton and Andrew Ryan and Ankit Ramchandani and Annie Dong and Annie Franco and Anuj Goyal and Aparajita Saraf and Arkabandhu Chowdhury and Ashley Gabriel and Ashwin Bharambe and Assaf Eisenman and Azadeh Yazdan and Beau James and Ben Maurer and Benjamin Leonhardi and Bernie Huang and Beth Loyd and Beto De Paola and Bhargavi Paranjape and Bing Liu and Bo Wu and Boyu Ni and Braden Hancock and Bram Wasti and Brandon Spence and Brani Stojkovic and Brian Gamido and Britt Montalvo and Carl Parker and Carly Burton and Catalina Mejia and Ce Liu and Changhan Wang and Changkyu Kim and Chao Zhou and Chester Hu and Ching-Hsiang Chu and Chris Cai and Chris Tindal and Christoph Feichtenhofer and Cynthia Gao and Damon Civin and Dana Beaty and Daniel Kreymer and Daniel Li and David Adkins and David Xu and Davide Testuggine and Delia David and Devi Parikh and Diana Liskovich and Didem Foss and Dingkang Wang and Duc Le and Dustin Holland and Edward Dowling and Eissa Jamil and Elaine Montgomery and Eleonora Presani and Emily Hahn and Emily Wood and Eric-Tuan Le and Erik Brinkman and Esteban Arcaute and Evan Dunbar and Evan Smothers and Fei Sun and Felix Kreuk and Feng Tian and Filippos Kokkinos and Firat Ozgenel and Francesco Caggioni and Frank Kanayet and Frank Seide and Gabriela Medina Florez and Gabriella Schwarz and Gada Badeer and Georgia Swee and Gil Halpern and Grant Herman and Grigory Sizov and Guangyi and Zhang and Guna Lakshminarayanan and Hakan Inan and Hamid Shojanazeri and Han Zou and Hannah Wang and Hanwen Zha and Haroun Habeeb and Harrison Rudolph and Helen Suk and Henry Aspegren and Hunter Goldman and Hongyuan Zhan and Ibrahim Damlaj and Igor Molybog and Igor Tufanov and Ilias Leontiadis and Irina-Elena Veliche and Itai Gat and Jake Weissman and James Geboski and James Kohli and Janice Lam and Japhet Asher and Jean-Baptiste Gaya and Jeff Marcus and Jeff Tang and Jennifer Chan and Jenny Zhen and Jeremy Reizenstein and Jeremy Teboul and Jessica Zhong and Jian Jin and Jingyi Yang and Joe Cummings and Jon Carvill and Jon Shepard and Jonathan McPhie and Jonathan Torres and Josh Ginsburg and Junjie Wang and Kai Wu and Kam Hou U and Karan Saxena and Kartikay Khandelwal and Katayoun Zand and Kathy Matosich and Kaushik Veeraraghavan and Kelly Michelena and Keqian Li and Kiran Jagadeesh and Kun Huang and Kunal Chawla and Kyle Huang and Lailin Chen and Lakshya Garg and Lavender A and Leandro Silva and Lee Bell and Lei Zhang and Liangpeng Guo and Licheng Yu and Liron Moshkovich and Luca Wehrstedt and Madian Khabsa and Manav Avalani and Manish Bhatt and Martynas Mankus and Matan Hasson and Matthew Lennie and Matthias Reso and Maxim Groshev and Maxim Naumov and Maya Lathi and Meghan Keneally and Miao Liu and Michael L. Seltzer and Michal Valko and Michelle Restrepo and Mihir Patel and Mik Vyatskov and Mikayel Samvelyan and Mike Clark and Mike Macey and Mike Wang and Miquel Jubert Hermoso and Mo Metanat and Mohammad Rastegari and Munish Bansal and Nandhini Santhanam and Natascha Parks and Natasha White and Navyata Bawa and Nayan Singhal and Nick Egebo and Nicolas Usunier and Nikhil Mehta and Nikolay Pavlovich Laptev and Ning Dong and Norman Cheng and Oleg Chernoguz and Olivia Hart and Omkar Salpekar and Ozlem Kalinli and Parkin Kent and Parth Parekh and Paul Saab and Pavan Balaji and Pedro Rittner and Philip Bontrager and Pierre Roux and Piotr Dollar and Polina Zvyagina and Prashant Ratanchandani and Pritish Yuvraj and Qian Liang and Rachad Alao and Rachel Rodriguez and Rafi Ayub and Raghotham Murthy and Raghu Nayani and Rahul Mitra and Rangaprabhu Parthasarathy and Raymond Li and Rebekkah Hogan and Robin Battey and Rocky Wang and Russ Howes and Ruty Rinott and Sachin Mehta and Sachin Siby and Sai Jayesh Bondu and Samyak Datta and Sara Chugh and Sara Hunt and Sargun Dhillon and Sasha Sidorov and Satadru Pan and Saurabh Mahajan and Saurabh Verma and Seiji Yamamoto and Sharadh Ramaswamy and Shaun Lindsay and Shaun Lindsay and Sheng Feng and Shenghao Lin and Shengxin Cindy Zha and Shishir Patil and Shiva Shankar and Shuqiang Zhang and Shuqiang Zhang and Sinong Wang and Sneha Agarwal and Soji Sajuyigbe and Soumith Chintala and Stephanie Max and Stephen Chen and Steve Kehoe and Steve Satterfield and Sudarshan Govindaprasad and Sumit Gupta and Summer Deng and Sungmin Cho and Sunny Virk and Suraj Subramanian and Sy Choudhury and Sydney Goldman and Tal Remez and Tamar Glaser and Tamara Best and Thilo Koehler and Thomas Robinson and Tianhe Li and Tianjun Zhang and Tim Matthews and Timothy Chou and Tzook Shaked and Varun Vontimitta and Victoria Ajayi and Victoria Montanez and Vijai Mohan and Vinay Satish Kumar and Vishal Mangla and Vlad Ionescu and Vlad Poenaru and Vlad Tiberiu Mihailescu and Vladimir Ivanov and Wei Li and Wenchen Wang and Wenwen Jiang and Wes Bouaziz and Will Constable and Xiaocheng Tang and Xiaojian Wu and Xiaolan Wang and Xilun Wu and Xinbo Gao and Yaniv Kleinman and Yanjun Chen and Ye Hu and Ye Jia and Ye Qi and Yenda Li and Yilin Zhang and Ying Zhang and Yossi Adi and Youngjin Nam and Yu and Wang and Yu Zhao and Yuchen Hao and Yundi Qian and Yunlu Li and Yuzi He and Zach Rait and Zachary DeVito and Zef Rosnbrick and Zhaoduo Wen and Zhenyu Yang and Zhiwei Zhao and Zhiyu Ma},
      year={2024},
      eprint={2407.21783},
      archivePrefix={arXiv},
      primaryClass={cs.AI},
      url={https://arxiv.org/abs/2407.21783}, 
}

@misc{nemotron4,
      title={Nemotron-4 340B Technical Report}, 
      author={Nvidia and : and Bo Adler and Niket Agarwal and Ashwath Aithal and Dong H. Anh and Pallab Bhattacharya and Annika Brundyn and Jared Casper and Bryan Catanzaro and Sharon Clay and Jonathan Cohen and Sirshak Das and Ayush Dattagupta and Olivier Delalleau and Leon Derczynski and Yi Dong and Daniel Egert and Ellie Evans and Aleksander Ficek and Denys Fridman and Shaona Ghosh and Boris Ginsburg and Igor Gitman and Tomasz Grzegorzek and Robert Hero and Jining Huang and Vibhu Jawa and Joseph Jennings and Aastha Jhunjhunwala and John Kamalu and Sadaf Khan and Oleksii Kuchaiev and Patrick LeGresley and Hui Li and Jiwei Liu and Zihan Liu and Eileen Long and Ameya Sunil Mahabaleshwarkar and Somshubra Majumdar and James Maki and Miguel Martinez and Maer Rodrigues de Melo and Ivan Moshkov and Deepak Narayanan and Sean Narenthiran and Jesus Navarro and Phong Nguyen and Osvald Nitski and Vahid Noroozi and Guruprasad Nutheti and Christopher Parisien and Jupinder Parmar and Mostofa Patwary and Krzysztof Pawelec and Wei Ping and Shrimai Prabhumoye and Rajarshi Roy and Trisha Saar and Vasanth Rao Naik Sabavat and Sanjeev Satheesh and Jane Polak Scowcroft and Jason Sewall and Pavel Shamis and Gerald Shen and Mohammad Shoeybi and Dave Sizer and Misha Smelyanskiy and Felipe Soares and Makesh Narsimhan Sreedhar and Dan Su and Sandeep Subramanian and Shengyang Sun and Shubham Toshniwal and Hao Wang and Zhilin Wang and Jiaxuan You and Jiaqi Zeng and Jimmy Zhang and Jing Zhang and Vivienne Zhang and Yian Zhang and Chen Zhu},
      year={2024},
      eprint={2406.11704},
      archivePrefix={arXiv},
      primaryClass={cs.CL},
      url={https://arxiv.org/abs/2406.11704}, 
}

@inproceedings{sglang,
author = {Zheng, Lianmin and Yin, Liangsheng and Xie, Zhiqiang and Sun, Chuyue and Huang, Jeff and Yu, Cody Hao and Cao, Shiyi and Kozyrakis, Christos and Stoica, Ion and Gonzalez, Joseph E. and Barrett, Clark and Sheng, Ying},
title = {SGLang: efficient execution of structured language model programs},
year = {2024},
isbn = {9798331314385},
publisher = {Curran Associates Inc.},
address = {Red Hook, NY, USA},
booktitle = {Proceedings of the 38th International Conference on Neural Information Processing Systems},
articleno = {2000},
numpages = {27},
location = {Vancouver, BC, Canada},
series = {NIPS '24}
}

@misc{polarsparsity,
      title={Polar Sparsity: High Throughput Batched LLM Inferencing with Scalable Contextual Sparsity}, 
      author={Susav Shrestha and Brad Settlemyer and Nikoli Dryden and Narasimha Reddy},
      year={2025},
      eprint={2505.14884},
      archivePrefix={arXiv},
      primaryClass={cs.LG},
      url={https://arxiv.org/abs/2505.14884}, 
}

@inproceedings{sarathiserve,
author = {Agrawal, Amey and Kedia, Nitin and Panwar, Ashish and Mohan, Jayashree and Kwatra, Nipun and Gulavani, Bhargav S. and Tumanov, Alexey and Ramjee, Ramachandran},
title = {Taming throughput-latency tradeoff in LLM inference with sarathi-serve},
year = {2024},
isbn = {978-1-939133-40-3},
publisher = {USENIX Association},
address = {USA},
booktitle = {Proceedings of the 18th USENIX Conference on Operating Systems Design and Implementation},
articleno = {7},
numpages = {18},
location = {Santa Clara, CA, USA},
series = {OSDI'24}
}

@article{pytorchdistributed,
author = {Li, Shen and Zhao, Yanli and Varma, Rohan and Salpekar, Omkar and Noordhuis, Pieter and Li, Teng and Paszke, Adam and Smith, Jeff and Vaughan, Brian and Damania, Pritam and Chintala, Soumith},
title = {PyTorch distributed: experiences on accelerating data parallel training},
year = {2020},
issue_date = {August 2020},
publisher = {VLDB Endowment},
volume = {13},
number = {12},
issn = {2150-8097},
url = {https://doi.org/10.14778/3415478.3415530},
doi = {10.14778/3415478.3415530},
journal = {Proc. VLDB Endow.},
month = aug,
pages = {3005–3018},
numpages = {14}
}

@misc{splitwise,
      title={Splitwise: Efficient generative LLM inference using phase splitting}, 
      author={Pratyush Patel and Esha Choukse and Chaojie Zhang and Aashaka Shah and Íñigo Goiri and Saeed Maleki and Ricardo Bianchini},
      year={2024},
      eprint={2311.18677},
      archivePrefix={arXiv},
      primaryClass={cs.AR},
      url={https://arxiv.org/abs/2311.18677}, 
}

@inbook{gpipe,
author = {Huang, Yanping and Cheng, Youlong and Bapna, Ankur and Firat, Orhan and Chen, Mia Xu and Chen, Dehao and Lee, HyoukJoong and Ngiam, Jiquan and Le, Quoc V. and Wu, Yonghui and Chen, Zhifeng},
title = {GPipe: efficient training of giant neural networks using pipeline parallelism},
year = {2019},
publisher = {Curran Associates Inc.},
address = {Red Hook, NY, USA},
booktitle = {Proceedings of the 33rd International Conference on Neural Information Processing Systems},
articleno = {10},
numpages = {10}
}

@inproceedings{pipedream,
author = {Narayanan, Deepak and Harlap, Aaron and Phanishayee, Amar and Seshadri, Vivek and Devanur, Nikhil R. and Ganger, Gregory R. and Gibbons, Phillip B. and Zaharia, Matei},
title = {PipeDream: generalized pipeline parallelism for DNN training},
year = {2019},
isbn = {9781450368735},
publisher = {Association for Computing Machinery},
address = {New York, NY, USA},
url = {https://doi.org/10.1145/3341301.3359646},
doi = {10.1145/3341301.3359646},
booktitle = {Proceedings of the 27th ACM Symposium on Operating Systems Principles},
pages = {1–15},
numpages = {15},
location = {Huntsville, Ontario, Canada},
series = {SOSP '19}
}

@misc{megatronlm,
      title={Megatron-LM: Training Multi-Billion Parameter Language Models Using Model Parallelism}, 
      author={Mohammad Shoeybi and Mostofa Patwary and Raul Puri and Patrick LeGresley and Jared Casper and Bryan Catanzaro},
      year={2020},
      eprint={1909.08053},
      archivePrefix={arXiv},
      primaryClass={cs.CL},
      url={https://arxiv.org/abs/1909.08053}, 
}

@inproceedings{GQA,
    title = "{GQA}: Training Generalized Multi-Query Transformer Models from Multi-Head Checkpoints",
    author = "Ainslie, Joshua  and
      Lee-Thorp, James  and
      de Jong, Michiel  and
      Zemlyanskiy, Yury  and
      Lebron, Federico  and
      Sanghai, Sumit",
    editor = "Bouamor, Houda  and
      Pino, Juan  and
      Bali, Kalika",
    booktitle = "Proceedings of the 2023 Conference on Empirical Methods in Natural Language Processing",
    month = dec,
    year = "2023",
    address = "Singapore",
    publisher = "Association for Computational Linguistics",
    url = "https://aclanthology.org/2023.emnlp-main.298/",
    doi = "10.18653/v1/2023.emnlp-main.298",
    pages = "4895--4901"
}

@misc{MQA,
      title={Fast Transformer Decoding: One Write-Head is All You Need}, 
      author={Noam Shazeer},
      year={2019},
      eprint={1911.02150},
      archivePrefix={arXiv},
      primaryClass={cs.NE},
      url={https://arxiv.org/abs/1911.02150}, 
}

@inproceedings{FlashAttention,
author = {Dao, Tri and Fu, Daniel Y. and Ermon, Stefano and Rudra, Atri and R\'{e}, Christopher},
title = {FLASHATTENTION: fast and memory-efficient exact attention with IO-awareness},
year = {2022},
isbn = {9781713871088},
publisher = {Curran Associates Inc.},
address = {Red Hook, NY, USA},
booktitle = {Proceedings of the 36th International Conference on Neural Information Processing Systems},
articleno = {1189},
numpages = {16},
location = {New Orleans, LA, USA},
series = {NIPS '22}
}

@misc{FlashAttention3,
      title={FlashAttention-3: Fast and Accurate Attention with Asynchrony and Low-precision}, 
      author={Jay Shah and Ganesh Bikshandi and Ying Zhang and Vijay Thakkar and Pradeep Ramani and Tri Dao},
      year={2024},
      eprint={2407.08608},
      archivePrefix={arXiv},
      primaryClass={cs.LG},
      url={https://arxiv.org/abs/2407.08608}, 
}

@inproceedings{ESPN,
author = {Shrestha, Susav and Reddy, Narasimha and Li, Zongwang},
title = {ESPN: Memory-Efficient Multi-vector Information Retrieval},
year = {2024},
isbn = {9798400706158},
publisher = {Association for Computing Machinery},
address = {New York, NY, USA},
url = {https://doi.org/10.1145/3652024.3665515},
doi = {10.1145/3652024.3665515},
booktitle = {Proceedings of the 2024 ACM SIGPLAN International Symposium on Memory Management},
pages = {95–107},
numpages = {13},
location = {Copenhagen, Denmark},
series = {ISMM 2024}
}

@article{espn+,
	author = {Shrestha, Susav and Gautam, Aayush and Reddy, Narasimha},
	date = {2025/03/13},
	doi = {10.1007/s11227-025-07118-9},
	id = {Shrestha2025},
	isbn = {1573-0484},
	journal = {The Journal of Supercomputing},
	number = {4},
	pages = {613},
	title = {Storage access optimization for efficient GPU-centric information retrieval},
	url = {https://doi.org/10.1007/s11227-025-07118-9},
	volume = {81},
	year = {2025}}

@misc{InfiniteLLM,
      title={Infinite-LLM: Efficient LLM Service for Long Context with DistAttention and Distributed KVCache}, 
      author={Bin Lin and Chen Zhang and Tao Peng and Hanyu Zhao and Wencong Xiao and Minmin Sun and Anmin Liu and Zhipeng Zhang and Lanbo Li and Xiafei Qiu and Shen Li and Zhigang Ji and Tao Xie and Yong Li and Wei Lin},
      year={2024},
      eprint={2401.02669},
      archivePrefix={arXiv},
      primaryClass={cs.DC},
      url={https://arxiv.org/abs/2401.02669}, 
}

@inproceedings{LongServe,
author = {Wu, Bingyang and Liu, Shengyu and Zhong, Yinmin and Sun, Peng and Liu, Xuanzhe and Jin, Xin},
title = {LoongServe: Efficiently Serving Long-Context Large Language Models with Elastic Sequence Parallelism},
year = {2024},
isbn = {9798400712517},
publisher = {Association for Computing Machinery},
address = {New York, NY, USA},
url = {https://doi.org/10.1145/3694715.3695948},
doi = {10.1145/3694715.3695948},
booktitle = {Proceedings of the ACM SIGOPS 30th Symposium on Operating Systems Principles},
pages = {640–654},
numpages = {15},
location = {Austin, TX, USA},
series = {SOSP '24}
}

@inproceedings{Speculativedecoding,
author = {Leviathan, Yaniv and Kalman, Matan and Matias, Yossi},
title = {Fast inference from transformers via speculative decoding},
year = {2023},
publisher = {JMLR.org},
booktitle = {Proceedings of the 40th International Conference on Machine Learning},
articleno = {795},
numpages = {13},
location = {Honolulu, Hawaii, USA},
series = {ICML'23}
}

@misc{Speculativesampling,
      title={Accelerating Large Language Model Decoding with Speculative Sampling}, 
      author={Charlie Chen and Sebastian Borgeaud and Geoffrey Irving and Jean-Baptiste Lespiau and Laurent Sifre and John Jumper},
      year={2023},
      eprint={2302.01318},
      archivePrefix={arXiv},
      primaryClass={cs.CL},
      url={https://arxiv.org/abs/2302.01318}, 
}

@misc{gammatune,
      title={Token-Driven GammaTune: Adaptive Calibration for Enhanced Speculative Decoding}, 
      author={Aayush Gautam and Susav Shrestha and Narasimha Reddy},
      year={2025},
      eprint={2504.00030},
      archivePrefix={arXiv},
      primaryClass={cs.CL},
      url={https://arxiv.org/abs/2504.00030}, 
}

@inproceedings{medusa,
author = {Cai, Tianle and Li, Yuhong and Geng, Zhengyang and Peng, Hongwu and Lee, Jason D. and Chen, Deming and Dao, Tri},
title = {MEDUSA: Simple LLM inference acceleration framework with multiple decoding heads},
year = {2024},
publisher = {JMLR.org},
booktitle = {Proceedings of the 41st International Conference on Machine Learning},
articleno = {203},
numpages = {27},
location = {Vienna, Austria},
series = {ICML'24}
}

@misc{eagle3,
      title={EAGLE-3: Scaling up Inference Acceleration of Large Language Models via Training-Time Test}, 
      author={Yuhui Li and Fangyun Wei and Chao Zhang and Hongyang Zhang},
      year={2025},
      eprint={2503.01840},
      archivePrefix={arXiv},
      primaryClass={cs.CL},
      url={https://arxiv.org/abs/2503.01840}, 
}

@misc{helixparallel,
      title={Helix Parallelism: Rethinking Sharding Strategies for Interactive Multi-Million-Token LLM Decoding}, 
      author={Nidhi Bhatia and Ankit More and Ritika Borkar and Tiyasa Mitra and Ramon Matas and Ritchie Zhao and Maximilian Golub and Dheevatsa Mudigere and Brian Pharris and Bita Darvish Rouhani},
      year={2025},
      eprint={2507.07120},
      archivePrefix={arXiv},
      primaryClass={cs.DC},
      url={https://arxiv.org/abs/2507.07120}, 
}

@misc{kvparallel,
      title={No Request Left Behind: Tackling Heterogeneity in Long-Context LLM Inference with Medha}, 
      author={Amey Agrawal and Haoran Qiu and Junda Chen and Íñigo Goiri and Chaojie Zhang and Rayyan Shahid and Ramachandran Ramjee and Alexey Tumanov and Esha Choukse},
      year={2025},
      eprint={2409.17264},
      archivePrefix={arXiv},
      primaryClass={cs.LG},
      url={https://arxiv.org/abs/2409.17264}, 
}

@misc{flashinfer,
      title={FlashInfer: Efficient and Customizable Attention Engine for LLM Inference Serving}, 
      author={Zihao Ye and Lequn Chen and Ruihang Lai and Wuwei Lin and Yineng Zhang and Stephanie Wang and Tianqi Chen and Baris Kasikci and Vinod Grover and Arvind Krishnamurthy and Luis Ceze},
      year={2025},
      eprint={2501.01005},
      archivePrefix={arXiv},
      primaryClass={cs.DC},
      url={https://arxiv.org/abs/2501.01005}, 
}

@misc{ringattention,
      title={Ring Attention with Blockwise Transformers for Near-Infinite Context}, 
      author={Hao Liu and Matei Zaharia and Pieter Abbeel},
      year={2023},
      eprint={2310.01889},
      archivePrefix={arXiv},
      primaryClass={cs.CL},
      url={https://arxiv.org/abs/2310.01889}, 
}

@inproceedings{hetis,
author = {Mo, Zizhao and Liao, Jianxiong and Xu, Huanle and Zhou, Zhi and Xu, ChengZhong},
title = {Hetis: Serving LLMs in Heterogeneous GPU Clusters with Fine-grained and Dynamic Parallelism},
year = {2025},
isbn = {9798400714665},
publisher = {Association for Computing Machinery},
address = {New York, NY, USA},
url = {https://doi.org/10.1145/3712285.3759784},
doi = {10.1145/3712285.3759784},
booktitle = {Proceedings of the International Conference for High Performance Computing, Networking, Storage and Analysis},
pages = {1710–1724},
numpages = {15},
location = {
},
series = {SC '25}
}

@misc{beyondbuzz,
      title={Beyond the Buzz: A Pragmatic Take on Inference Disaggregation}, 
      author={Tiyasa Mitra and Ritika Borkar and Nidhi Bhatia and Ramon Matas and Shivam Raj and Dheevatsa Mudigere and Ritchie Zhao and Maximilian Golub and Arpan Dutta and Sailaja Madduri and Dharmesh Jani and Brian Pharris and Bita Darvish Rouhani},
      year={2025},
      eprint={2506.05508},
      archivePrefix={arXiv},
      primaryClass={cs.DC},
      url={https://arxiv.org/abs/2506.05508}, 
}

@article{shuffleinfer,
author = {Hu, Cunchen and Huang, Heyang and Xu, Liangliang and Chen, Xusheng and Wang, Chenxi and Xu, Jiang and Chen, Shuang and Feng, Hao and Wang, Sa and Bao, Yungang and Sun, Ninghui and Shan, Yizhou},
title = {ShuffleInfer: Disaggregate LLM Inference for Mixed Downstream Workloads},
year = {2025},
issue_date = {June 2025},
publisher = {Association for Computing Machinery},
address = {New York, NY, USA},
volume = {22},
number = {2},
issn = {1544-3566},
url = {https://doi.org/10.1145/3732941},
doi = {10.1145/3732941},
journal = {ACM Trans. Archit. Code Optim.},
month = jul,
articleno = {77},
numpages = {24}
}

@misc{kernelflume,
    title={KernelFlume: Elastic Core-Attention Scaling for Agentic Long-Context Decoding},
    author={Xiang, Guangyu and Kang, Xueze and Zhang, Lin and Lin, Wenxiang and Shi, Shaohuai and Wang, Yuxin and Chu, Xiaowen},
    year={2026},
    eprint={2606.29207},
    archivePrefix={arXiv},
    primaryClass={cs.DC},
    url={https://arxiv.org/abs/2606.29207}
}

@misc{ruler,
      title={RULER: What's the Real Context Size of Your Long-Context Language Models?}, 
      author={Cheng-Ping Hsieh and Simeng Sun and Samuel Kriman and Shantanu Acharya and Dima Rekesh and Fei Jia and Yang Zhang and Boris Ginsburg},
      year={2024},
      eprint={2404.06654},
      archivePrefix={arXiv},
      primaryClass={cs.CL},
      url={https://arxiv.org/abs/2404.06654}, 
}

@misc{leval,
 title={L-Eval: Instituting Standardized Evaluation for Long Context Language Models},
 author={An, Chenxin and Gong, Shansan and Zhong, Ming and Zhao, Xingjian and Li, Mukai and Zhang, Jun and Kong, Lingpeng and Qiu, Xipeng},
 year={2023},
 eprint={2307.11088},
 archivePrefix={arXiv},
 url={https://arxiv.org/abs/2307.11088}
}

@misc{longbench,
 title={LongBench: A Bilingual, Multitask Benchmark for Long Context Understanding},
 author={Bai, Yushi and Lv, Xin and Zhang, Jiajie and Lyu, Hongchang and Tang, Jiankai and Huang, Zhidian and Du, Zhengxiao and Liu, Xiao and Zeng, Aohan and Hou, Lei and Dong, Yuxiao and Tang, Jie and Li, Juanzi},
 year={2023},
 eprint={2308.14508},
 archivePrefix={arXiv},
 url={https://arxiv.org/abs/2308.14508}
}

@misc{longbenchv2,
 title={LongBench v2: Towards Deeper Understanding and Reasoning on Realistic Long-context Multitasks},
 author={Bai, Yushi and Tu, Shangqing and Zhang, Jiajie and Peng, Hao and Wang, Xiaozhi and Lv, Xin and Cao, Shulin and Xu, Jiazheng and Hou, Lei and Dong, Yuxiao and Tang, Jie and Li, Juanzi},
 year={2024},
 eprint={2412.15204},
 archivePrefix={arXiv},
 url={https://arxiv.org/abs/2412.15204}
}

@misc{sglang2024dpattention,
  author       = {{SGLang}},
  title        = {{SGLang} v0.4: Zero-Overhead Batch Scheduler, Cache-Aware Load Balancer, Faster Structured Outputs},
  year         = {2024},
  month        = dec,
  howpublished = {\url{https://lmsys.org/blog/2024-12-04-sglang-v0-4/}},
  note         = {LMSYS Org Blog. Accessed October 2026}
}
